\documentclass[12]{article}
\usepackage{graphicx}
\begin{document}
\title{Target-ionization with exchange in Ps-atom scattering}
\author{{ Hasi Ray } \\ 
{\it Institute of Plasma Research, Bhat, Gandhinagar, Ahmedabad 382428, India } \\
E-mail:    hasi\_ray@yahoo.com \& hasiray@ipr.res.in}
\date{}
\maketitle
\vskip 0.05cm   
{\bf Abstract:} A proposal is made by the University College of London Group [1] for measuring the target-ionization cross sections in Ps and atom scattering. We calculate the corresponding theoretical data for Ps-H and Ps-He systems including the effect of exchange on Coulomb-Born approximation for the projectile-elastic and projectile-inelastic processes and report the total target-ionization cross sections for the first time. \\

{\bf PACS No.} 36.10.Dr, 34.70.+e, 82.30.Gg 

\vskip 0.5cm   
The exotic atoms like positronium (Ps), protonium etc. are interesting probes to investigate new physics due to the fact that they are formed by a matter and its corresponding antimatter. Their charge and mass centers coincide. The unique properties of Ps promise to yield stimulating findings [1-35].
Ionization is an important phenomena in Ps and atom scattering [1,2,14-16,18,19,22-33]. Only Ps break-up is studied extensively using reliable methods [15,16,26-28], but no emphasis was given on target ionization [17]. Recently Laricchia et al [1] proposed measurements on target-ionization with and without projectile ionization, but no reliable theoretical data are available for
the estimation.
 Due to zero first order polarization potential, the effect of exchange plays a predominant role in Ps-atomic system.
We include exchange to study target-ionization for Ps-H and Ps-He scattering using Coulomb-Born approximation. In my knowledge this is the first reliable calculation to estimate cross sections for target ionization without projectile ionization. 
Earlier calculations [14,29-33] on target-ionization did not include the effect of exchange. \\

We choose the post form of the scattering amplitude, $\langle \Psi_f\mid v_f\mid\Psi_i^+\rangle$ (following the conventional notation); however in direct channel both the post and prior forms are equal. We treat the moving Ps as a plane wave in the incident channel as well as in the final channel; the target ionized electron is represented by a continuum Coulomb wave which was a bound atomic electron in the incident channel. Being a neutral system, the plane wave approximation for the moving Ps may not introduce an error. In addition the first order effect due to polarizability vanishes in such a system of Ps and atom. So the present theory is quite reliable and it is reflected in the comparison of first-Born approximation (FBA) and the close coupling approximation (CCA) results of the earlier calculation [34,35]. Only at very low energies, the FBA and CCA elastic cross sections differ in Ps-H [34] and Ps-Li [35] systems. \\
 
Inclusion of exchange is a rigorous and tedious job. The direct scattering amplitudes using FBA and CBA vanish if Ps does not change the parity. We are interested in projectile-elastic, [Ps(1s)$\rightarrow$Ps(1s)] and projectile-inelastic, [Ps(1s)$\rightarrow$Ps(2s)] target-ionization channels since they should have the maximum contribution than other target-ionization channels. For both the processes in both the systems, the entire contribution to target-ionization cross section is coming only from the exchange channels since the direct matrix elements vanish [14,29-32]. The exchange amplitude for Ps-He system is expressed as\\

$$G_{\bf k}^{He}(\hat{\bf k}_f) = -\frac{1}{\pi}\int e^{-i{\bf k_f.R_2}}\eta_{1s}^*(\mbox{\boldmath$\rho$}_2)\Phi_f^*\{{\bf r_1, r_3}\}[V_{He}^G]e^{i{\bf k_i.R_1}}\eta_{1s}(\mbox{\boldmath$\rho$}_1)\Phi_i\{{\bf r_2, r_3}\}d{\bf x}d{\bf r_1}d{\bf r_2}d{\bf r_3} \eqno(1) $$
with
$$ V_{He}^G = \frac{Z}{\mid{\bf x}\mid}-\frac{Z}{\mid{\bf r_2}\mid}-\frac{1}{\mid{{\bf x}-{\bf r_1}}\mid} +\frac{1}{\mid{{\bf r_2}-{\bf r_1}}\mid}-\frac{1}{\mid{{\bf x}-{\bf r_3}}\mid}+\frac{1}{\mid{{\bf r_2}-{\bf r_3}}\mid} \eqno(2) $$
\\
with
   ${\bf R}_j = \frac{1}{2}({\bf x}+{\bf r}_j)$
     and
$\mbox{\boldmath$\rho$}_j = ({\bf x} - {\bf r}_j)$; j=1,2. Here,
${\bf x}$ is the coordinate of positron in Ps, and
${\bf r}_j$; j = 1 to 3 are those of electrons in Ps and He respectively in the incident channel with respect to the center of mass of the system. Functions
$\eta$ and $\Phi$ indicate the wave functions of Ps and He respectively. Subscript `$i$` identifies the incident channel, whereas `$f$` represents the final channel. Accordingly ${\bf k}_i$ and ${\bf k}_f$ are the momenta of the projectile in the initial and final channels respectively. Z is the nuclear charge of the target helium atom.
The Ps wave function is considered at ground state in both the incident and final channels. If we remove the third electron from the Ps-He system which is represented by the position coordinate ${\bf r}_3$, the expression (1) should fit to Ps-H system.\\
 The triply differential cross section (TDCS) for the break-up of He in Ps-He scattering is defined as

$$ \frac{d^3\sigma(E_i)}{d\hat{\bf k_f}d\hat{\bf k}dE_{\bf k}} = \frac{k_f k}{k_i}\{\mid G_{\bf k}^{He} \mid^2 \} \eqno(3) $$
and the integrated cross section as\\
$${\sigma(E_i)} = \int d\hat{\bf k}_f \int d\hat{\bf k}\int dE_{\bf k}
     \frac{d^3\sigma(E_i)}{d\hat{\bf k_f}d\hat{\bf k}dE_{\bf k}} \eqno(4)
$$ \\
with $dE_{\bf k} = k dk$, ${\bf k}$ is the momentum of the ionized electron. 
The ground state Ps wave function is $\eta_{1s}(\rho) = e^{-\rho/2}/\sqrt{8\pi}$. The target helium wave function in the incident channel is $\phi_i({\bf r}_2,{\bf r}_3) = u_{1s}({\bf r}_2) u_{1s}({\bf r}_3)$, and in final channel is $\phi_f({\bf r}_1,{\bf r}_3) = u_{\bf k}^*({\bf r}_1) u_{1s}({\bf r}_3)$, 
where $u_{1s}{(\bf r)} = \lambda^{3/2}\pi^{-1/2}e^{-\lambda r}$ with $\lambda = 1.6875$ [36]; $u_{\bf k}({\bf r})$ is the continuum Coulomb wave function with momentum $\bf k$ and is expressed as
 $$ u_{\bf k}({\bf r}) = (2\pi)^{-3/2}e^{-\gamma\pi/2}
\Gamma(1-i\gamma)e^{i{\bf k}.{\bf r}} {_1F_1}[i\gamma,1,-i(kr+{\bf k}.{\bf r})]  \eqno(5)$$
We choose $\gamma=-\lambda/k$ so that $u_{\bf k}({\bf r})$ is orthogonal to $u_{1s}({\bf r})$ following Klar et al [see ref. 36].
\\

After carrying on the integration over ${\bf r}_3$, the equation (1) transformed to \\
$$G_{\bf k}^{He}(\hat{\bf k}_f) = - \frac{\lambda_i^{3/2} e^{-\gamma\pi/2} \Gamma(1 + i \gamma)}{16\sqrt{2}\pi^4} \{\sum_{i=1}^4 I_i + 8 \frac{{(\lambda_i\lambda_f)}^{3/2}}{\lambda_{if}^3} \sum_{i=5}^6 I_i\} \eqno(6)$$
with
$$I_1 = \int d{\bf x}d{\bf r}_1 d{\bf r}_2 e^{i\frac{1}{2}{\bf Q}.{\bf x}}e^{i(\frac{1}{2}{\bf k}_i - {\bf k}).{\bf r}_1}e^{-i\frac{1}{2}{\bf k}_f.{\bf r}_2} 
\frac{1}{\mid{\bf x}\mid} e^{-\lambda_ir_2} e^{-\alpha_i\mid{\bf x}-{\bf r}_1\mid} e^{-\alpha_f\mid{\bf x}-{\bf r}_2\mid} {_1F_1}[-i\gamma,1,i(kr_1+{\bf k}.{\bf r}_1)] \eqno(7a) $$ 
$$I_2 = -\int d{\bf x}d{\bf r}_1 d{\bf r}_2 e^{i\frac{1}{2}{\bf Q}.{\bf x}}e^{i(\frac{1}{2}{\bf k}_i - {\bf k}).{\bf r}_1}e^{-i\frac{1}{2}{\bf k}_f.{\bf r}_2} 
\frac{1}{\mid{{\bf r}_2}\mid} e^{-\lambda_ir_2} e^{-\alpha_i\mid{\bf x}-{\bf r}_1\mid} e^{-\alpha_f\mid{\bf x}-{\bf r}_2\mid} {_1F_1}[-i\gamma,1,i(kr_1+{\bf k}.{\bf r}_1)] \eqno(7b)$$ 
$$I_3 = -\int d{\bf x}d{\bf r}_1 d{\bf r}_2 e^{i\frac{1}{2}{\bf Q}.{\bf x}}e^{i(\frac{1}{2}{\bf k}_i - {\bf k}).{\bf r}_1}e^{-i\frac{1}{2}{\bf k}_f.{\bf r}_2} 
\frac{1}{\mid{{\bf x}-{\bf r}_1}\mid} e^{-\lambda_ir_2} e^{-\alpha_i\mid{\bf x}-{\bf r}_1\mid} e^{-\alpha_f\mid{\bf x}-{\bf r}_2\mid} {_1F_1}[-i\gamma,1,i(kr_1+{\bf k}.{\bf r}_1)] \eqno(7c) $$ 
$$I_4 = \int d{\bf x}d{\bf r}_1 d{\bf r}_2 e^{i\frac{1}{2}{\bf Q}.{\bf x}}e^{i(\frac{1}{2}{\bf k}_i - {\bf k}).{\bf r}_1}e^{-i\frac{1}{2}{\bf k}_f.{\bf r}_2} 
\frac{1}{\mid{{\bf r}_2-{\bf r}_1}\mid} e^{-\lambda_ir_2} e^{-\alpha_i\mid{\bf x}-{\bf r}_1\mid} e^{-\alpha_f\mid{\bf x}-{\bf r}_2\mid} {_1F_1}[-i\gamma,1,i(kr_1+{\bf k}.{\bf r}_1)] \eqno(7d) $$ 
$$I_5 =  \{ I(\lambda_{if}=0) -  I  + \frac{1}{2}\lambda_{if}(\frac{\partial}{\partial\lambda_{if}})I \} \eqno(8a)$$ 
$$I_6 =  \{ I^\prime(\lambda_{if}=0)  -  I^\prime  + \frac{1}{2}\lambda_{if}(\frac{\partial}{\partial\lambda_{if}})I^\prime \} \eqno(8b)$$
and
$$I = -\int d{\bf x}d{\bf r}_1 d{\bf r}_2 e^{i\frac{1}{2}{\bf Q}.{\bf x}}e^{i(\frac{1}{2}{\bf k}_i - {\bf k}).{\bf r}_1}e^{-i\frac{1}{2}{\bf k}_f.{\bf r}_2} 
\frac{e^{-\lambda_{if} X}}{X} e^{-\lambda_ir_2} e^{-\alpha_i\mid{\bf x}-{\bf r}_1\mid} e^{-\alpha_f\mid{\bf x}-{\bf r}_2\mid} {_1F_1}[-i\gamma,1,i(kr_1+{\bf k}.{\bf r}_1)] \eqno(8c) $$ 
$$I^\prime = \int d{\bf x}d{\bf r}_1 d{\bf r}_2 e^{i\frac{1}{2}{\bf Q}.{\bf x}}e^{i(\frac{1}{2}{\bf k}_i - {\bf k}).{\bf r}_1}e^{-i\frac{1}{2}{\bf k}_f.{\bf r}_2} 
\frac{e^{-\lambda_{if} r_2}}{r_2} e^{-\lambda_ir_2} e^{-\alpha_i\mid{\bf x}-{\bf r}_1\mid} e^{-\alpha_f\mid{\bf x}-{\bf r}_2\mid} {_1F_1}[-i\gamma,1,i(kr_1+{\bf k}.{\bf r}_1)] \eqno(8d) $$ 
Here $\alpha_i$, $\alpha_f$ are the screening parameters used in the wave function of Ps and $\lambda_i$, $\lambda_f$ are the screeing parameters used to define
 wave function of the target electron, in the incident and final channels 
respectively;$\lambda_{if}=\lambda_i+\lambda_f$.

The Ps-H system contains only the first four integrals $I_1$, $I_2$, $I_3$, $I_4$. The terms like $I_5$ and $I_6$ appear in all the multi-electron targets. We derive the following standard integrals which are useful to reduce the dimension of the above six integrals:

$$   \int d{\bf r} e^{i{\bf k}.{\bf r}} e^{-\beta r} e^{-\alpha\mid{\bf x}-{\bf r}\mid} = 2\pi\alpha\beta \int_0^1 dy y(1-y) e^{iy{\bf k}.{\bf x}}(\frac{1}{\mu}\frac{\partial}{\partial\mu})^2\frac{1}{\mu}e^{-\mu x}   \eqno(A)        $$
when $\mu^2=y\beta^2+(1-y)\alpha^2+y(1-y)k^2$;

$$  \int d{\bf r} e^{i{\bf k}.{\bf r}} \frac{e^{-\beta r}}{r} e^{-\alpha\mid{\bf x}-{\bf r}\mid} = -2\pi\alpha \int_0^1 dy y e^{i(1-y){\bf k}.{\bf x}}(\frac{1}{\mu}\frac{\partial}{\partial\mu})\frac{1}{\mu}e^{-\mu x} \eqno(B)   $$
when $\mu^2=y\alpha^2+(1-y)\beta^2+y(1-y)k^2$;

$$  \int d{\bf r} e^{i{\bf k}.{\bf r}} e^{-\beta r} \frac{e^{-\alpha\mid{\bf x}-{\bf r}\mid}}{\mid{\bf x}-{\bf r}\mid} = -2\pi\beta \int_0^1 dy y e^{iy{\bf k}.{\bf x}}(\frac{1}{\mu}\frac{\partial}{\partial\mu})\frac{1}{\mu}e^{-\mu x}  \eqno(C) $$
when $\mu^2=y\beta^2+(1-y)\alpha^2+y(1-y)k^2$;

$$  \int d{\bf r} e^{i{\bf k}.{\bf r}} e^{-\alpha_1\mid{\bf x}_1-{\bf r}\mid} e^{-\alpha_2\mid{\bf x}_2-{\bf r}\mid} = 2\pi\alpha_1\alpha_2 \int_0^1 dy y(1-y) e^{i(1-y)n{\bf k}.{\bf x}_2} e^{iy{\bf k}.{\bf x}_1}(\frac{1}{\mu}\frac{\partial}{\partial\mu})^2\frac{1}{\mu}e^{-\mu\mid{\bf x}_1-{\bf x}_2\mid}  \eqno(D) $$
when $\mu^2=y\alpha_2^2+(1-y)\alpha_1^2+y(1-y)k^2$;

$$  \int \frac{e^{-\lambda r}}{\mid{\bf x}-{\bf r}\mid} d{\bf r} = \frac{8\pi}{\lambda^3}  [\frac{1}{x} - \frac{e^{-\lambda x}}{x} - \frac{1}{2}\lambda e^{-\lambda x}] \eqno(E) $$

In addition, we use the help of another standard integral

$$   \int\frac{e^{-\beta r}}{r} e^{i{\bf P}.{\bf r}}{_1F_1[-i\gamma,1,i(kr+{\bf k}.{\bf r})]i} d{\bf r} = \frac{4\pi}{T} \{\frac{({\bf P}+{\bf k})^2-(k+i\beta)^2}{T}\}^{i\gamma}   \eqno(F) $$
where $T = \beta^2 + {\bf P}^2$. After simplification each of the six integrals in equations (7) and (8), transforms into a two-dimensional form. Lastly, the numerical integrations are performed using the Gauss-Legendre quadrature method. The Gauss-Legendre quadratures are generated using our own code.\\

We introduce the integrated cross section for the projectile elastic [Ps(1s)$\rightarrow$ Ps(1s)] and projectile inelastic [Ps(1s)$\rightarrow$ Ps(2s)] target-ionization for the Ps-H and Ps-He scattering. In figure 1, we presented the results of Ps-H system and in figure 2, the Ps-He system. We compared them with the corresponding target-elastic Ps-ionization cross sections [26,27]. For both the processes the direct scattering amplitude vanishes, so only the exchange contributes. Our earlier target-ionization cross sections using FBA [33] and CBA [30,31] include the contribution of all the important direct channels, but the present channels had null contribution due to parity conservation of Ps. The results reported in ref.[30] is superior than earlier [31,33] due to the fact that the target continuum Coulomb wave function was orthogonal to the target atomic wave function. We see that the present projectile-inelastic [Ps(1s)$\rightarrow$Ps(2s)] target-ionization cross sections have a negligible contribution in comparison to the projectile-elastic [Ps(1s)$\rightarrow$Ps(1s)] target-ionization cross sections in both the systems. So we can expect that the contribution of exchange is not so important in projectile inelastic target-ionization channels and inclusion of only the direct channels are sufficient. So we adapt the CBA results for the rest projectile-inelastic target-ionization channels. The present summed target-ionization cross section is comparable with the summed direct target-ionization cross section [30,32] at lower incident energies. At higher incident energies, the present summed target-ionization cross section is negligible in comparison to summed direct target-ionization cross section [30,32]. This is in complete agreement with the basic theory of exchange. To get total target-ionization cross section we add the present summed cross section with the previous [30,32] projectile-inelastic target-ionization cross sections using CBA and presented in figure 3 for Ps-H and figure 4 for Ps-He systems. We follow the similar methodology for Ps-ionization, consider the effect of exchange as negligible for the target-inelastic Ps-ionization processes and adapt the CBA results for target-inelastic Ps-ionization. The summed total Ps-ionization cross sections are plotted in the same figures 3 and 4. We conclude that target-ionization processes are more important in the Ps-H system than the Ps-He system and projectile ionization is more important in the Ps-He system than the Ps-H system. These are in complete agreement with our earlier findings [30-32].  We present our present tabular data and summed CBA results for projectile-inelastic target ionization in table 1 for future investigations. \\

\begin{figure*}
\centering
\includegraphics[width=0.95\columnwidth]{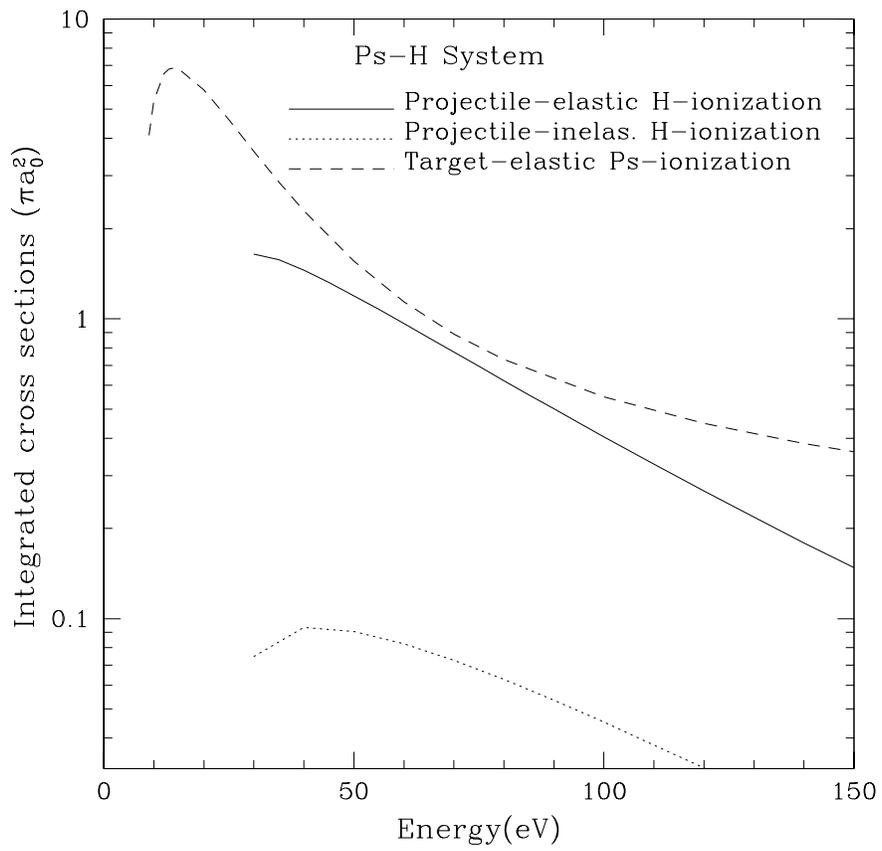}
\caption{
Integrated ionization cross sections in $\pi a_0^2$ for Ps-H scattering.}%
\end{figure*}
\begin{figure*}
\centering
\includegraphics[width=0.95\columnwidth]{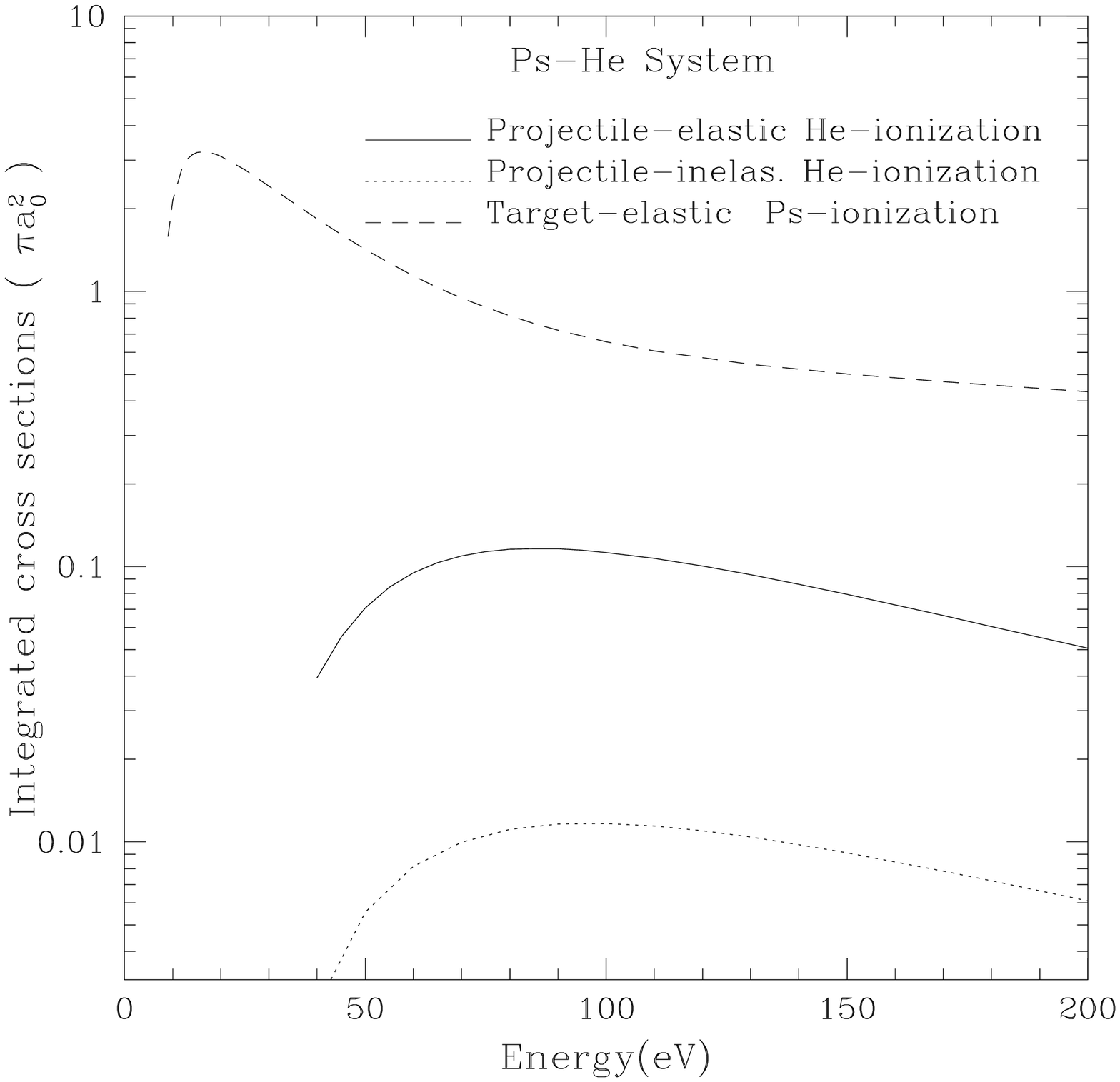}
\caption{
Integrated ionization cross sections in $\pi a_0^2$ for Ps-He scattering.}%
\end{figure*}
\begin{figure*}
\centering
\includegraphics[width=0.95\columnwidth]{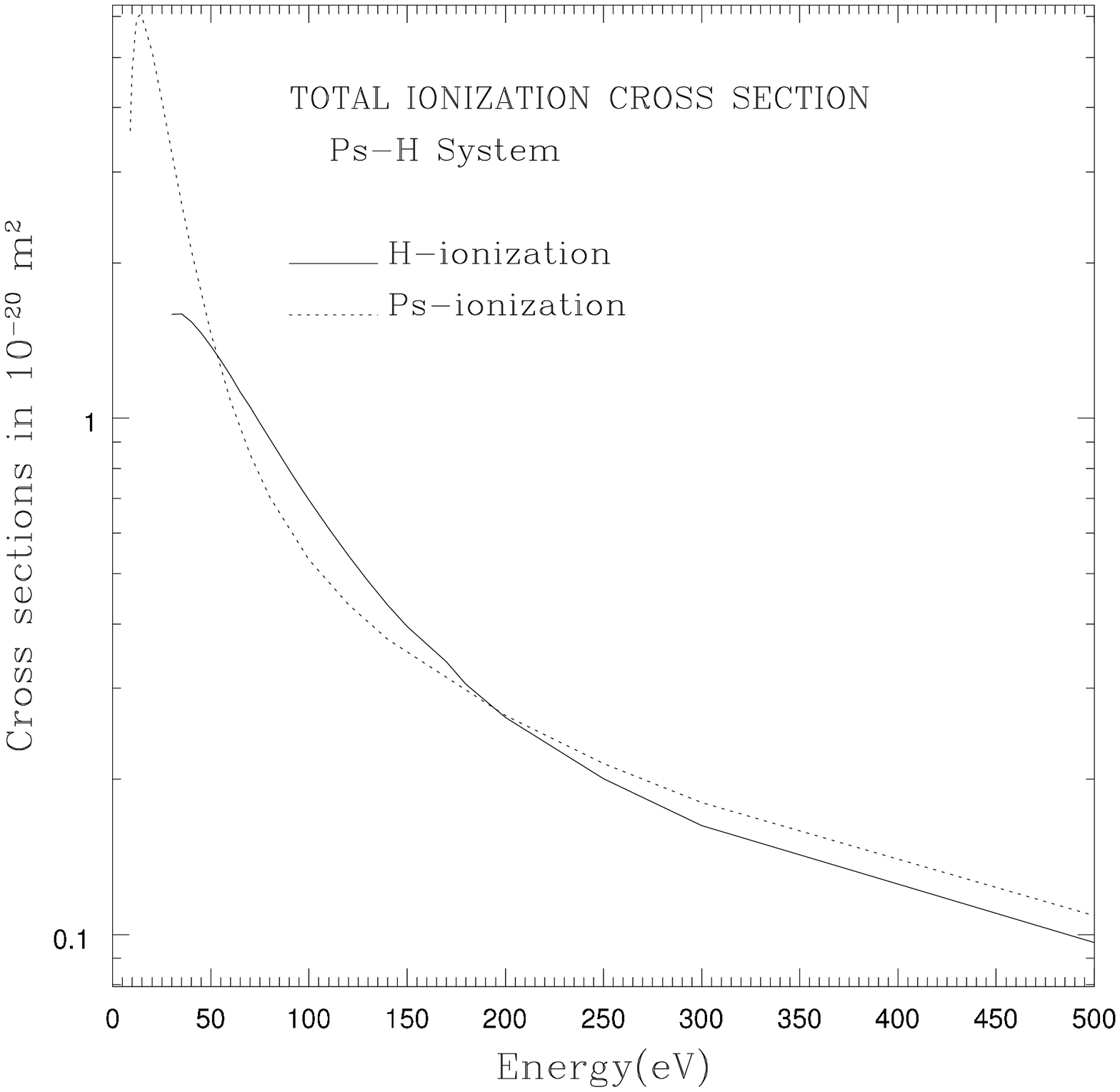}
\caption{
Total ionization cross sections in $10^{-20}m^2$ for Ps-H scattering.}
\end{figure*}
\begin{figure*}
\centering
\includegraphics[width=0.95\columnwidth]{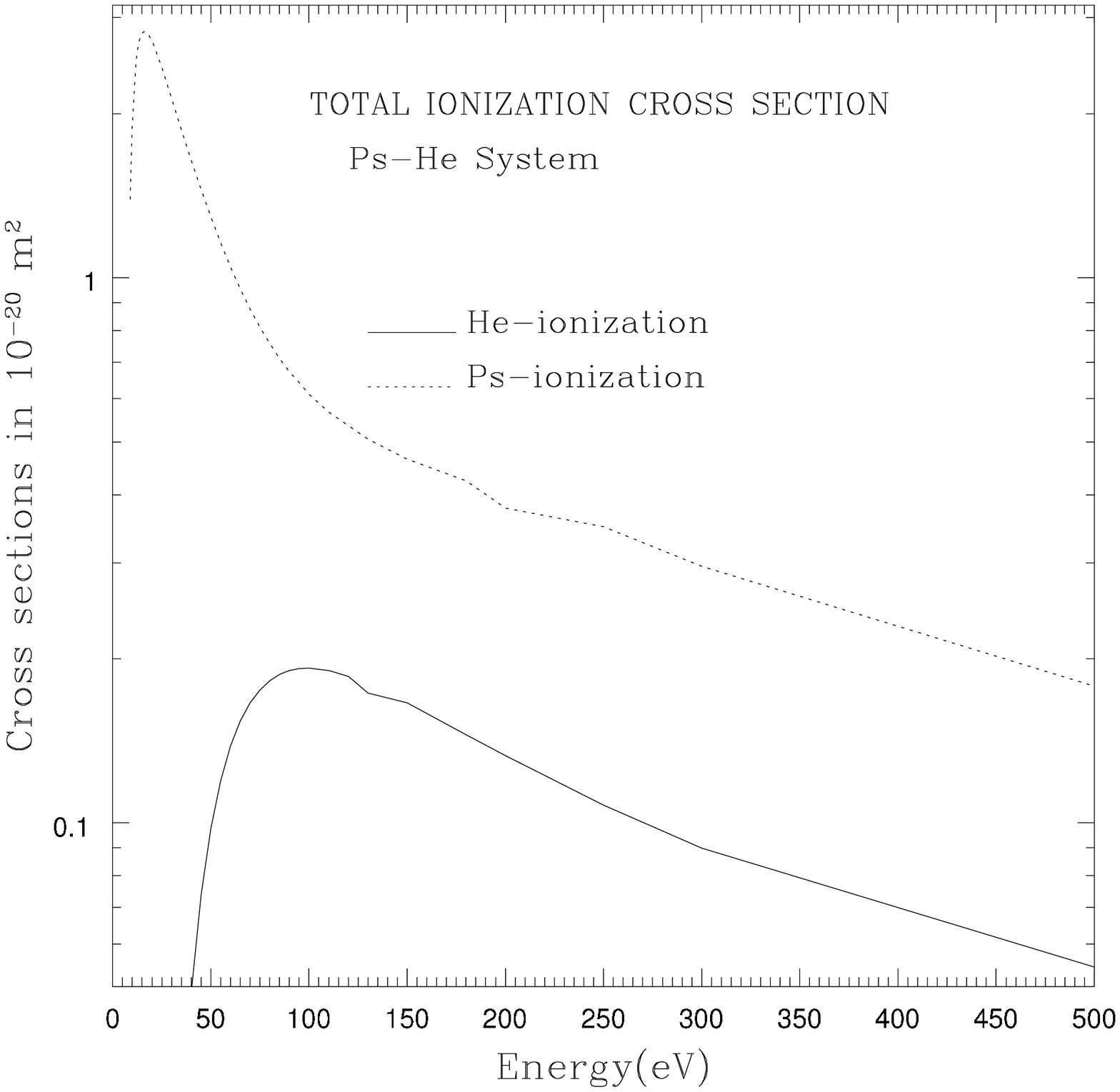}
\caption{
Total ionization cross sections in $10^{-20}m^2$ for Ps-He scattering.}
\end{figure*}
\begin{table*}[h]
\begin{center}
Table 1. Target-ionization cross sections for Ps-H and Ps-He scattering.\\
\begin{tabular}{|c|c|c|c|c|c|c|}
\hline
  &\multicolumn{6}{|c|}{Integrated target-ionization cross sections in $10^{-20}m^2$}\\
\cline{2-7}
Energy&\multicolumn{3}{|c|}{Ps-H system}&\multicolumn{3}{|c|}{Ps-He system} \\
\cline{2-4} \cline{5-7}
(eV) &\multicolumn{1}{|c|}{Elastic} &\multicolumn{2}{|c|}{Excitation} &\multicolumn{1}{|c|}{Elastic} & \multicolumn{2}{|c|}{Excitation} \\
\cline{3-4} \cline{6-7}
     &\multicolumn{1}{|c|}{Ps(1s)$\rightarrow$ Ps(1s)} &\multicolumn{1}{|c|}{Ps(1s)$\rightarrow$ Ps(2s)}&\multicolumn{1}{|c|}{Other}&\multicolumn{1}{|c|}{Ps(1s)$\rightarrow$ Ps(1s)} &\multicolumn{1}{|c|}{Ps(1s)$\rightarrow$ Ps(2s)}&\multicolumn{1}{|c|}{Other}\\
\hline
        30.0   & 1.4462  &   0.0657 & 0.0762 &           &         &\\
        35.0   & 1.3839  &   0.0783 & 0.1302 &           &         &   \\
        40.0   & 1.2782  &   0.0823 & 0.1791 &   0.0347  &    0.0022&0.0123  \\
        45.0   & 1.1627  &   0.0821 & 0.2193 &   0.0490  &    0.0036&0.0213  \\
        50.0   & 1.0500  &   0.0798 & 0.2503 &   0.0623  &    0.0049&0.0307  \\
        55.0   & 0.9446  &   0.0765 & 0.2733 &   0.0739  &    0.0061&0.0396   \\
        60.0   & 0.8481  &   0.0726 &0.2894 &   0.0834  &    0.0072&0.0478  \\
        65.0   & 0.7605  &   0.0683 & 0.3002 &   0.0907  &    0.0081&0.0550  \\
        70.0   & 0.6816  &   0.0640 & 0.3068 &   0.0960  &    0.0088&0.0612  \\
        75.0   & 0.6109  &   0.0595 & 0.3101 &   0.0995  &   0.0093&0.0666   \\
        80.0   & 0.5475  &   0.0552 & 0.3110 &   0.1015  &   0.0098&0.0711   \\
        85.0   & 0.4910  &   0.0510 & 0.3100 &   0.1023  &   0.0100&0.0749   \\
        90.0   & 0.4405  &   0.0470 & 0.3076 &   0.1019  &   0.0102&0.0781   \\
        95.0   & 0.3956  &   0.0433 & 0.3042 &   0.1008  &   0.0103&0.0807  \\
       100.0   & 0.3555  &   0.0397 & 0.3002 &   0.0991  &   0.0102&0.0828  \\
       110.0   & 0.2880  &   0.0334 & 0.2906 &   0.0943  &   0.0100&0.0859  \\
       120.0   & 0.2344  &   0.0280 & 0.2801 &   0.0884  &   0.0096&0.0877  \\
       130.0   & 0.1917  &   0.0234 & 0.2694 &   0.0821  &   0.0092&0.0885  \\
       140.0   & 0.1575  &   0.0196 &  &   0.0758  &   0.0086 & \\
       150.0   & 0.1301  &   0.0165 & 0.2485 &   0.0696  &   0.0080&0.0884  \\
       160.0   & 0.1079  &   0.0138 &  &   0.0638  &   0.0074 & \\
       170.0   & 0.0900  &   0.0117 &  &   0.0583  &   0.0069 & \\
       180.0   & 0.0754  &   0.0098 & 0.2207 &   0.0533  &   0.0063&0.0857  \\
       200.0   & 0.0536  &   0.0071 & 0.2048 &   0.0444  &   0.0054&0.0831  \\
       220.0   & 0.0388  &   0.0052 & &   0.0370  &   0.0045&  \\
       250.0   & 0.0246  &   0.0033 & 0.1726 &   0.0283  &   0.0035&0.0760  \\
       300.0   & 0.0124  &   0.0017 & 0.1487 &   0.0184  &   0.0023&0.0692  \\
       400.0   & 0.0038  &   0.0005 &  &   0.0084  &   0.0011 & \\
       500.0   & 0.0015  &   0.0002 & 0.0950 &   0.0042  &   0.0005&0.0497  \\
\hline
\end{tabular}
\end{center}
\end{table*} %

To conclude, we report reliable theoretical data for total target-ionization cross sections in Ps-H and Ps-He scattering including the effect of exchange on the most important projectile-elastic and the projectile inelastic channels.
\\

{\bf Acknowledgement}

Author is thankful to Institute of Plasma Research, Bhat, Gandhinagar for the visiting scientist position.
\\

{\bf References:} 

1. Private communication with G. Larcchia, UCL (2006).

2. H. S. W. Massey and C. B. O. Mohr, Proc. Phys. Soc. {\bf 67}, 695 (1954).

3. M. I. Barker and B. H. Bransden, J. Phys. B {\bf 1}, 1109 (1968).

4. J. Madox, Nature {\bf 314}, 315 (1985).

5. J. R. Manson and R. H. Ritche, Phys. Rev. Lett. {\bf 54}, 785 (1985).

6. C. K. Au and R. J. Drachman, Phys. Rev. Lett. {\bf 56}, 324 (1986).

7. S. K. Adhikari, Phys. Lett. A {\bf 294}, 308 (2002).

8. D. M. Schrader, Phys. Rev. Lett. {\bf 92}, 043401 (2004).

9. J. DiRienzi and R. J. Drachman, Phys. Rev. A {\bf 76}, 032705 (2007).

10. H. Ray, Phys. Rev. A {\bf 73}, 064501 (2006).

11. C. M. Surko, Nature {\bf 449}, 153 (2007).

12. D. B. Cassidy and A. P. Mills Jr., Nature {\bf 449}, 195 (2007).

13. N. Patel, Nature {\bf 449}, 133 (2007).

14. M. T. McAlinden, F. G. R. S. MacDonald and H. R. J.Walters, Can. J. Phys., {\bf 74}, 434 (1996).

15. H. R. J. Walters, C. Starrett and M. T. McAlinden, NIMB, {\bf 247}, 111 (2006).

16. C. P. Campbell, M. T. McAlinden, F. G. R. S. MacDonald and H. R. J. Walters, Phys. Rev. Lett. {\bf 80}, 5097 (1998).

17. J. E. Blackwood, M. T. McAlinden and H. R. J. Walters, Phys. Rev. A {\bf 65},030502(R) (2002).

18. S. Amritage, D. E. Leslie, J. Beale, G. Laricchia, NIMB, {\bf 247}, 98 (2006).

19. S. Armitage, D. E. Leslie, A. J. Garner and G. Laricchia, Phys. Rev. Lett {\bf 89}, 173402 (2002).

20. N. Suzuki, T. Hirade, F. Saito and T. Hyodo, Rad. Phys. \& Chem. {\bf 68}, 647 (2003).

21. M. Skalsey, J. J. Engbrecht, R. K. Bithell, R. S. Vallery and D. W. Gidley, Phys. Rev. Lett. {\bf 80}, 3727 (1998).

22. H. Ray, NASA GSFC Science Symposium on Atomic \& Molecular Physics, edited by A.K.Bhatia, p. 121 (2007).

23. H. R. J. Walters and C. Starrett, NASA GSFC Science Symposium on Atomic \& Molecular Physics, edited by A.K.Bhatia, p. 187 (2007).

24. L. Sarkadi, Phys. Rev. A {\bf 68}, 032706 (2003).

25. P. K. Biswas and Sadhan K. Adhikari, Phys. Rev. A {\bf 59}, 363 (1999).

26. H. Ray, Euro. Phys. Lett. {\bf 73}, 21 (2006).

27. H. Ray, PRAMANA {\bf 66}, 415 (2006).

28. Singly differential cross section for Ps break-up using CBOA, reported by Hasi Ray in XXV-ICPEAC (2007).

29. H. Ray, PRAMANA {\bf 63}, 1063 (2004).

30. H. Ray, J.Phys.B {\bf 35}, 3365 (2002).

31. H. Ray, NIMB {\bf 192}, 191 (2002).

32. H. Ray, Phys. Lett. A {\bf 299}, 65 (2002).

33. H. Ray, Phys. Lett.A {\bf 252}, 316 (1999).

34. H. Ray and A. S. Ghosh, J. Phys. B. {\bf 31}, 4427 (1998).

35. H. Ray, J. Phys. B {\bf 33}, 4285 (2000).

36. H. Ray and A. C. Roy, J. Phys. B {\bf 22}, 1425 (1989).

\end{document}